\documentclass[aps,prl,twocolumn,groupedaddress,showpacs,floatfix,superscriptaddress,longbibliography]{revtex4-2}
\usepackage{epsfig}
\usepackage{amsmath,amssymb}
\usepackage{graphicx}
\usepackage[dvipsnames,usenames]{xcolor}
\usepackage[normalem]{ulem}
\usepackage{soul}  
\usepackage[hyperindex,breaklinks,colorlinks=true, citecolor = blue]{hyperref}
\emergencystretch=\maxdimen
\begin{document}

\title{Phase Diagram of the Su-Schrieffer-Heeger-Hubbard model on a square lattice}
\author{Chunhan Feng}
\affiliation{Department of Physics, University of California, Davis, CA 95616,USA}
\author{Bo Xing}
\affiliation{Science, Mathematics and Technology Cluster, Singapore University of Technology and Design, 8 Somapah Road, 487372 Singapore}
\author{Dario Poletti}
\affiliation{Science, Mathematics and Technology Cluster, Singapore University of Technology and Design, 8 Somapah Road, 487372 Singapore}
\affiliation{Engineering Product Development Pillar, Singapore University of Technology and Design, 8 Somapah Road, 487372 Singapore}
\author{Richard Scalettar}
\affiliation{Department of Physics, University of California, Davis, CA 95616,USA}
\author{George Batrouni}
\affiliation{Universit\'e C\^ote d'Azur, CNRS, Institut de Physique de Nice (INPHYNI), 06000 Nice, France}
\affiliation{Centre for Quantum Technologies, National University of Singapore; 2 Science Drive 3 Singapore 117542}
\affiliation{Department of Physics, National University of Singapore, 2 Science Drive 3, 117542 Singapore} 
\affiliation{Beijing Computational Science Research Center, Beijing 100193, China}

\date{\today}

\begin{abstract}
The Hubbard and Su-Schrieffer-Heeger Hamiltonians (SSH) are iconic models
for understanding the qualitative effects
of electron-electron and electron-phonon interactions respectively. In the
two-dimensional square lattice Hubbard model at
 half filling, the on-site Coulomb repulsion, $U$, between up and down electrons
 induces antiferromagnetic (AF)
 order and a Mott insulating phase.
On the other hand, for the SSH model, there is an AF phase when the
 electron-phonon coupling $\lambda$ is less than a 
critical value $\lambda_c$ and a bond order wave when $\lambda > \lambda_c$.
 In this work, we
 perform numerical studies on the square lattice optical Su-Schrieffer-Heeger-Hubbard
 Hamiltonian (SSHH),
 which combines both interactions. We use the determinant quantum Monte Carlo
 (DQMC) method which does
 not suffer from the fermionic sign problem at half filling. We map out the phase
 diagram and find that it exhibits a direct first-order transition
 between an antiferromagnetic phase and a bond-ordered wave as $\lambda$ increases. 
The AF phase is characterized by two different regions. At smaller $\lambda$ the
 behavior is similar to that of the pure Hubbard model; the other region, while maintaining
 long range AF order, exhibits larger kinetic energies and double occupancy, {\it
 i.e.}~larger quantum fluctuations, similar to the AF phase found in the pure SSH model. 

\end{abstract}

\maketitle

{\it Introduction:} 
Electron-electron and electron-phonon interactions play important roles in 
determining the ground state properties of many-body systems. Over the past
 decades, much computational  effort has been put into studying systems that
 feature one or the other of these interactions. One of the most widely used
 models to study the effect of electron-electron interaction with on-site
 repulsion $U$ is the Hubbard model~\cite{Hubbard1963} which exhibits
 metallic, ferromagnetic, antiferromagnetic (AF) and superconducting (SC)
 orders, as well as intricate inhomogeneous spin and charge patterns, depending
 on $U$ and the doping \cite{georges96, LeeWen2006}.
 The physics of the square lattice  Hubbard model bears remarkable resemblance
 to that of the cuprate superconductors.
Two of the most commonly studied electron-phonon Hamiltonians are the Holstein
 \cite{holstein59} and the Su-Schrieffer-Heeger (SSH) \cite{su79} models. Their
 fundamental difference is that in the former, electrons and phonons interact
 on a single site, while in the latter, the electron-phonon interactions occur
 on the bonds, {\it i.e.~}in the tunneling
 term. The Holstein interaction is widely used to explore polaron and charge-density wave (CDW) physics \cite{Noack1991,Vekic1992,Zhang2019,CChen2019,Feng2020,Weber2018,Cohen-Stead2019,ZXLi2019,Feng2020i,Nosarzewski2021,Xiao2021,Bradley2021}, and conventional $s$-wave SC
 transitions\cite{dee2020relative,Nosarzewski2021},
while the SSH interaction occurs in systems like conjugate polymers \cite{keiss92}, organic
 charge transfer salts \cite{ishiguro90}, metal salts \cite{toftlund84} and CuGeO$_3$
 \cite{hase93}. 

In the two-dimensional square lattice, the half-filled Holstein model predicts the
 emergence of a
CDW phase at any value of the electron-phonon interaction $\lambda$ \cite{scalettar89}.
 In the presence of an additional on-site electron-electron repulsion $U$, 
 the system can exhibit dominant AF or CDW correlations depending on the relative
 magnitude of $U$
 and $\lambda$ \cite{koller2004first,werner2007efficient}. Interestingly, there are
 indications of an intermediate metallic
 phase between the AF and CDW phases
 \cite{Fehske_2008,JohnstonDeveraux2013, WangDemler2020, costa2020},
 as well as other exotic regimes\cite{han2020strong}. 

For the 2D square lattice SSH model at half-filling, it was shown~\cite{XingBatrouni2021}
 that a finite critical electron-phonon interaction, $\lambda_c$, is needed to
 establish the bond-order-wave (BOW) phase, and 
weak antiferromagnetism was detected \cite{CaiYao2021, GoetzAssaad2021}  for
 $\lambda < \lambda_c$ despite the absence of $U$.  In the dilute limit, where
 bipolarons are expected to condense into a superfluid at very high temperatures,
 AF is revealed as well in the effective Hamiltonian \cite{johnsousmonaberciu2018}.
 The cause of this antiferromagnetism is that, on a given bond, only electrons of
 different spin can tunnel simultaneously, resulting in a lowering of the energy 
via the electron-phonon coupling on the bonds and
an {\it increase } in the magnitude of the kinetic energy. In contrast, in the 
Hubbard model at half-filling, AF order emerges in a two-step process in which 
$U$ first  suppresses doubly occupied sites, and then AF order occurs due to a
 small remnant exchange process $J \sim 4t^2/U$.  The AF phase in the Hubbard
 limit is thereby accompanied by low kinetic energy. This distinction will play a
 role in a cross-over behavior we observe in the Su-Schrieffer-Heeger-Hubbard 
(SSHH) phase diagram.

\begin{figure}[htpb]
    \centering
    \includegraphics[width=\columnwidth]{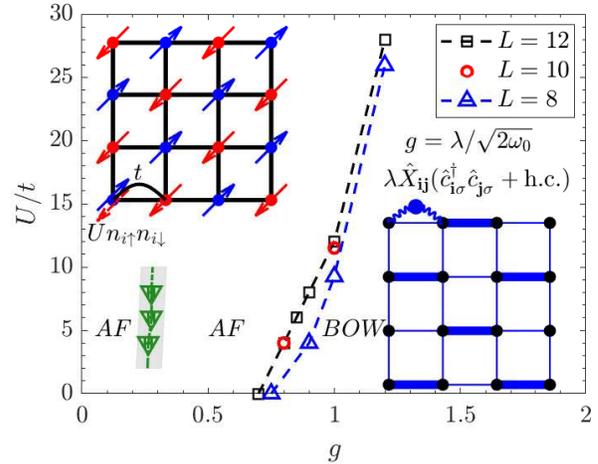}
    \caption{Phase diagram of the SSHH model at half-filling. $g$ is the dimensionless
 electron-phonon coupling constant, and $U/t$ is the Coulomb repulsion strength.
 A dotted (green) line shows the location of a crossover in the nature of the AF.   
$\beta = 16$ ensures the system is close to the ground state for all three lattice
 sizes. The AF-BOW transitions for $L=10,12$ coincide, indicating
 negligible finite size effects. The insets show schematically the AF and BOW phases.}
    \label{fig:PD}
\end{figure}

We study here the rich interplay of BOW and AF regimes in the SSHH model. Crucially, 
since the phonons couple to the electrons via the kinetic term, particle-hole symmetry
 is preserved and there is no sign problem (SP) at half-filling. This allows us to
 use determinant quantum Monte Carlo (DQMC) to study systems up to $12\times 12$ in 
size and at very low temperature. This contrasts with the Hubbard-Holstein model,
 where the SP precludes crossing the CDW-AF phase boundary\cite{costa2020}. Our 
resulting phase diagram  (Fig.~\ref{fig:PD}) exposes phases of long range  AF
 and BOW order. Prior to our work, only the quantum critical point along the
 $U=0$ axis (the SSH Hamiltonian) had been determined\cite{XingBatrouni2021}.
 A central observation of this work is that there are, within the AF phase,
 distinct regimes at small and intermediate electron-phonon coupling $\lambda$.
 The AF structure factor, double occupancy  and kinetic energy remain almost constant
 for small $\lambda$.  
However, for larger $\lambda$ these quantities show a marked dependence on $\lambda$. 
 As a consequence,  we will argue that the competition between $\lambda$ and $U$
 results not only in the expected AF-BOW transition, but also in a novel crossover
 {\it within} the AF phase. This crossover is clearly signaled in the AF correlations
 themselves, and also in the double occupancy, kinetic energy and pairing structure
 factors. 
These changes result from competition of the localizing effect of the Hubbard term
 and the quantum fluctuations favored by the SSH term, although they both can lead to AF. 


{\it Model and method:} We study the square lattice optical SSHH model, where the
 electronic hopping is modulated by an electron-phonon interaction and an on-site Coulomb 
repulsion is present. The Hamiltonian is 
\begin{align} 
    \mathcal{H} =& - t \sum_{\langle i, j \rangle, \sigma} 
\left( 1 - \lambda \hat{X}_{ij} \right)
  \left( \hat{c}^{\dagger}_{i \sigma} \hat{c}^{\phantom\dagger}_{j \sigma}
 + {\mathrm{h.c.}} \right)
  - \mu \sum_{i, \sigma} \hat{n}_{i \sigma} \nonumber \\
                 & + \sum_{\langle i, j\rangle} \left(\frac{1}{2M} \hat{P}_{ij}^2 +
  \frac{M}{2} \omega_{0}^2 \hat{X}_{ij}^2 \right) \nonumber \\
                 & + U \sum_{i} \left(\hat{n}_{i\uparrow}-\frac{1}{2}\right)
  \left(\hat{n}_{i\downarrow}-\frac{1}{2}\right)
  \label{eq:ham},
\end{align}
where $\hat{c}_{i \sigma}$ ($\hat{c}_{i \sigma}^{\dagger}$) destroys (creates) an electron of spin $\sigma=\uparrow,\downarrow$ on site $i$, $\mu$ is the electron chemical potential, $M$
 is the phonon mass and $\omega_{0}$ the oscillation frequency. The bond phonon
 displacement operator $\hat{X}_{ij}$ connects nearest neighbor sites
 $\langle i, j \rangle$; its conjugate bond momentum is $\hat{P}_{ij}$. 
 In the following, the magnitude of electron phonon coupling
 is given by the dimensionless parameter $g=\lambda  /\sqrt{2M\omega_{0}/\hbar}$,
so that the coupling term is
$ t g  (\hat b^{\phantom\dagger}_{ij}+\hat b^{\dagger}_{ij})
  ( \hat{c}^{\dagger}_{i \sigma}
  \hat{c}^{\phantom\dagger}_{j \sigma} + {\mathrm{h.c.}} ) $.
The on-site Coulomb repulsion is $U/t$, and $\hat{n}_{i\sigma}= 
\hat{c}^{\dagger}_{i \sigma} \hat{c}^{\phantom\dagger}_{i \sigma}$
 is the number operator on site $i$.  We work in units for which 
$\hbar = t = M = 1$ and fix $\omega_{0} = 1$.

The Hubbard-Stratonovich (HS) transformation is used in DQMC~\cite{Blankenbecler81, scalettar89, noack91, batrouni19a}, to express the
 quartic Coulomb interaction in quadratic form \cite{Stratonovich1957, Hubbard1959}.  The fermions are integrated
 out, yielding a determinant of a matrix that has dimension of the number of
 spatial sites $N$. The entries of the matrix depend on the HS
 and phonon fields.  We focus on half-filling ($\mu=0$), which does not present a SP,
 and work with $\beta = L_{\tau} \Delta \tau = 16$, where $L_{\tau} \sim 320$ is
 the number of imaginary slices, and $\Delta \tau$ is the imaginary time step.
 This $\beta$ is sufficiently large to access the ground state of the SSH model
 on the lattice sizes under investigation here~\cite{XingBatrouni2021}.

To characterize the emerging phases, we calculate the average kinetic energy in the $x$
  and $y$ directions, $\langle K_{x \left( y \right)} \rangle =
  \langle \hat{c}_{i,\sigma}^{\dagger} 
\hat{c}^{\phantom{\dagger}}_{i + \hat{x} \left( \hat{y} \right),\sigma} + h.c. \rangle  $,
  and the average phonon displacement in the $x$ and $y$ directions,
 $\langle X_{x \left( y \right)} \rangle $.
  These give insight into the broken $x$-$y$ symmetry in the BOW phase.
We also study the antiferromagnetic, $\langle S_{i}^{x} S_{i + r}^{x} \rangle $,
 and the bond order
  correlation functions, $\langle K_{x \left( y \right)} \left( i \right)
 K_{x \left( y \right)} \left( i + r \right)\rangle $.
Their Fourier transforms, $S_{\mathrm{AF}}$ and $S_{K_{x \left( y \right)}}$,
 are respectively the AF and BOW structure factors.  
In addition, we examine the double occupancy, $D = \langle \hat{n}_{i\uparrow}
 \hat{n}_{i\downarrow} \rangle$ and the total kinetic energy $\langle K \rangle =
 \langle K_x \rangle + \langle K_y \rangle$ which provide additional important insight.

{\it Results:} It is well known\cite{hirsch1989antiferromagnetism,white89a} that, at
 half filling, the square lattice Hubbard model, Eq.(\ref{eq:ham}) with $\lambda=0$,
 exhibits an AF phase for any $U>0$. Similarly, it was recently
 established\cite{CaiYao2021, GoetzAssaad2021} that the two-dimensional SSH model,
 Eq.(\ref{eq:ham}) with $U=0$, exhibits, at low temperature, an AF phase for small
 $\lambda$ and a BOW\cite{XingBatrouni2021} when $\lambda$ exceeds a critical value.
  Here we address the unknown structure of the phase diagram in the $(g,U/t)$ plane. 
\begin{figure}[htpb]
    \centering
    \includegraphics[width=\columnwidth]{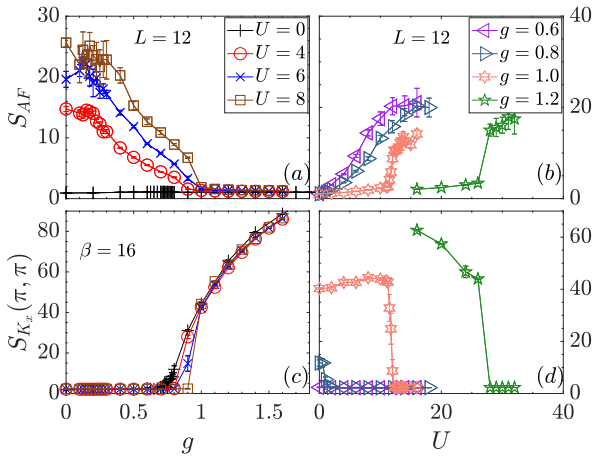}
    \caption{DQMC results of the AF (BOW) structure factor $S_{\mathrm{AF}}$
 ($S_{K_{x}}\left( \pi, \pi \right)$) for horizontal (left) and vertical (right)
 cuts in the phase diagram. In the AF phase, $S_{\mathrm{AF}}$ is finite and
 $S_{K_{x}}\left( \pi, \pi \right)$ is negligible. In the BOW phase,
 $S_{\mathrm{AF}}$ is negligible and $S_{K_{x}}\left( \pi, \pi \right)$ is finite. 
    }
    \label{fig:SF}
\end{figure}

To this end, we determine the phase boundaries with vertical and horizontal cuts,
 {\it i.e.}~by fixing $g$ ($U/t$) and studying the behavior of the system as
 $U/t$ ($g$) is changed. The AF and BOW phases are revealed by their respective
 structure factors, $S_{\mathrm{AF}}$ and $S_{K_{x}} \left( \pi, \pi \right)$.
 For low temperature and large systems, we start simulations with a phonon
 configuration that favors the BOW phase in the $x$ direction
 (bottom right inset Fig.~\ref{fig:PD}) because this structure is found to
 melt rapidly in the AF phase, but takes a long equilibration time to form.
 We measure $S_{K_{x \left( y \right)}} \left( k_{x}, k_{y} \right)$ for all
 momenta and observe a peak only at $S_{K_{x}} \left( \pi, \pi \right)$ when
 the system is in the BOW phase (see more details in \cite{supp}).
 Comparison of data for $L=10, 12$ indicate negligible finite size effects.
 
Figure~\ref{fig:SF}(a,c) shows the structure factors versus the
 dimensionless $g$ for several fixed values of $U/t$. For $U/t=4,6,8$ the
 system is a Hubbard AF for $g=0$, and remains AF as $g$ increases up to a critical
 value, $g_c(U/t)$. For $g<g_c$, $S_{K_{x}}$ is small, indicating the absence of BOW.
 $S_{K_{x}}$ then rises rapidly upon entry into the BOW phase at $g>g_c$. 

The behavior of the AF structure factor  $S_{\mathrm{AF}}$ is more subtle.  It is
 nonzero for  $g<g_c$, but there is an appreciable change in behavior well
 {\it before} its value drops precipitously: $S_{\mathrm{AF}}$ 
is initially constant for small $g$, Fig.~\ref{fig:SF}(a), but starts decreasing
 at $g_*\approx 0.2$. A finite size scaling analysis~\cite{supp} shows that AF
 regions exhibiting true long range order on both sides of $g_*$. The difference
 between these two AF regions, inferred from the kinetic energy and double occupancy
 will be discussed below. Indeed, since data for structure factors are more noisy 
than local correlation functions, these complementary observables will present
 additional compelling evidence for the crossover behavior at $g_*$.
  
Returning to the AF-BOW transition with increasing $g$, we see, Fig.~\ref{fig:SF}(a),
 that when $S_{\mathrm{AF}}$ drops, $S_{K_{x}}$ becomes nonzero.
 This occurs at $g_c \sim 1$ for $L = 12$. 
In Fig.~\ref{fig:SF}(b,d) we show the same quantities as panels (a,c) but now $g$
 is fixed and $U/t$ varies. For $g=0.6$, $S_{K_{x}}$ is small for all $U/t$ while
 $S_{\mathrm{AF}}$ increases smoothly as $U/t$ increases. For this value of $g$
 the system is always AF. For $g \geq 0.8$, $S_{\mathrm{AF}}$ is very small
 (essentially zero) while $S_{K_{x}}$ is large up to a $g$-dependent critical value,
 $U_c(g)$, indicating that the system is in the BOW phase~\cite{XingBatrouni2021}.
 At $U_c(g)$, there is a first order transition from the BOW to the AF phase,
 with clear discontinuous jumps in the order parameters.
This first order character is also observed 
for the larger $U$ values in the horizontal cuts (sweeping $g$ at fixed $U$) in
 Fig.~\ref{fig:SF}(a,c).

\begin{figure}[htpb]
    \centering
    \includegraphics[width=\columnwidth]{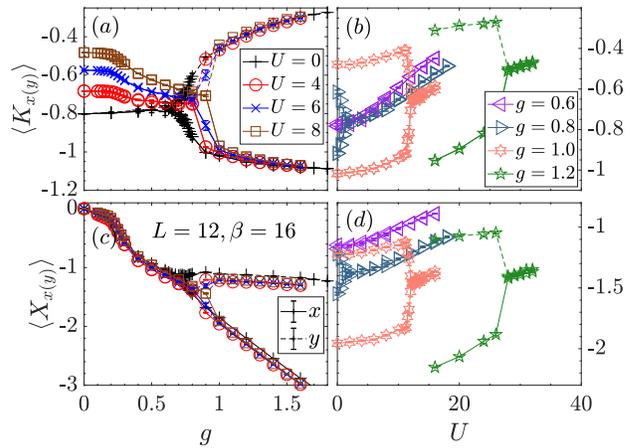}
    \caption{
    DQMC results of average kinetic energy, $\langle K_{x\left( y \right)} \rangle $,
 and average phonon displacement, $\langle X_{x \left( y \right)} \rangle$, in $x$
 and $y$ directions for horizontal (left) and vertical (right) cuts in the
 phase diagram.}
    \label{fig:KE_X}
\end{figure}

As shown in Ref.~\cite{XingBatrouni2021}, the BOW has $(\pi,\pi)$ ordering vector
 either in $x$ or in $y$ with two sublattice possibilities in each direction,
 resulting in the $Z_4$ symmetry breaking (in the thermodynamic limit). We now
 focus on this symmetry breaking as the system leaves the AF phase and enters
 the BOW phase. In the AF phase, the average kinetic energy and phonon displacement
 in the $x$ and $y$ directions are equal. In the BOW phase, the average
 kinetic energy and phonon displacement which align with the BOW direction
 increase in magnitude. We show in Fig.~\ref{fig:KE_X} the behavior of these
 quantities for the same parameters as in Fig.~\ref{fig:SF}. In panels (a,c),
 the $x$-$y$ symmetry is preserved in the AF phase, $g<g_c(U)$, and broken
 immediately when the system enters the BOW phase. This is clearly seen in
 the bifurcation in $K_{x \left( y \right)}$ and $X_{x \left( y \right)}$
 at $g_c$. As the on-site interaction becomes stronger, the electron-phonon
 coupling strength required to establish the BOW phase becomes larger.
 
In Fig.~\ref{fig:KE_X}(b,d), for constant $g$, the $x$-$y$ symmetry is broken
 for $U<U_c(g)$, $g \geq 0.8$ and restored immediately when the system exits
 the BOW and enters the AF phase at $U_{c}(g)$. For $g = 0.6$, the system is
 never in the BOW phase for all $U$ and therefore the $x$-$y$ symmetry is always
 preserved. The values of $g_c(U)$ and $U_c(g)$ obtained in Fig.~\ref{fig:SF} and
 Fig.~\ref{fig:KE_X} are in close agreement. We remark that, as observed in both
 figures, a small increase in $g$ (i.e. from $1.0$ to $1.2$) leads to significant changes
 of $U_{c}$ (i.e. $12$ to $26$). Putting these cuts at constant $g$ and $U$ 
together yields the phase diagram shown in Fig.~\ref{fig:PD}.

\begin{figure}[htpb]
    \centering
     \includegraphics[width=\columnwidth]{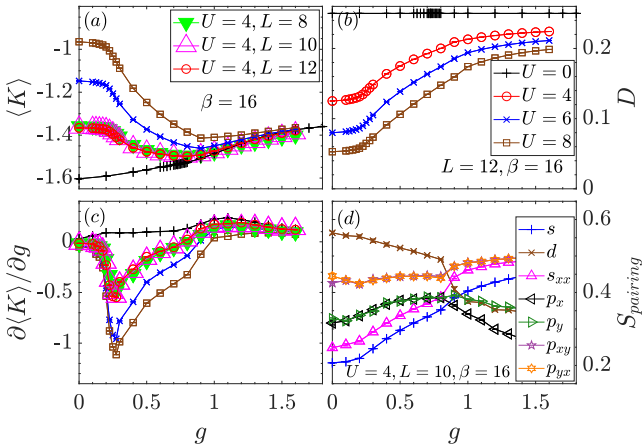}
    \caption{
    (a) Average kinetic energy. (b) double occupancy for
    different $U$ and fixed lattice size $L=12$;  (c) derivative
 of the kinetic energy with respect to $g$. The legends in panels (a,b) explain the symbols in panels (a,b,c). (d) pairing structure factors at fixed $U=4$ and $L=10$.
 }
    \label{fig:D_KE_Spairings}
\end{figure}

We now focus on the two AF regions
(separated by the vertical dotted line in Fig.~\ref{fig:PD}) for which
 $S_{\mathrm{AF}}$ provided initial evidence. We recall that for $g=0$,
 the system is in the Hubbard AF phase for any $U>0$, while for $U=0$,
 the system is in the SSH AF phase\cite{CaiYao2021,GoetzAssaad2021} for
 small $g$. The SSH AF at $U=0$ clearly has a different mechanism from the
 traditional two step Hubbard model process of moment formation at energy 
scale $U$ followed by moment ordering at energy scale $J \sim 4t^2/U$. A close
 analysis of Fig.~\ref{fig:KE_X}(a,c) shows that both $\langle K_{x(y)}\rangle$
 and $\langle X_{x(y)}\rangle$ remain almost constant for $g\lesssim 0.2$ and
 then increase in magnitude for $g> 0.2$. Similarly, the AF structure factor in
 Fig.~\ref{fig:SF}(a) is approximately constant for $g\lesssim 0,2$ and decreases for
 larger values of $g$. 

In Fig.~\ref{fig:D_KE_Spairings}(a), we show the average kinetic energy as a
 function of $g$ for several values of $U$.  $\langle K \rangle $ clearly 
exhibits a change of behavior at $g_*\approx 0.2$, supporting what is seen in
 Fig.~\ref{fig:SF} for $S_{\rm AF}$.  This is captured even more effectively
 in Fig.~\ref{fig:D_KE_Spairings}(c), which shows a sharp peak at $g \sim 0.275$ 
in $\partial \langle K \rangle/ \partial g$ vs $g$. A comparison between 
 $\langle K \rangle$ given in this SSHH Hamiltonian and in an (approximate)
 `effective' Hubbard model \cite{supp} gives more insight on this crossover.
 Figure \ref{fig:D_KE_Spairings}(b) shows the double occupancy, $D$, which
 increases in value for $g\gtrsim 0.2$. This behavior (larger kinetic energy 
and double occupancy) indicate that the system has left the ``large U'' Hubbard AF,
 where both quantities are suppressed, and entered an AF region strongly 
influenced by the SSH electron-phonon coupling, where quantum fluctuations are
 large. Going from one of these AF regions to the other is a crossover,
 not a phase transition. 
 Nevertheless, there is a clear signature in the
 increased quantum fluctuations.

Since the pure SSH Hamiltonian preserves $O(4)$ symmetry, and an AF/CDW/SC
 degenerate ground state is expected in the anti-adiabatic
 limit~\cite{CaiYao2021,GoetzAssaad2021}, it is useful to examine the
 superconducting  structure factor $S_{\rm pairing}$,
the spatial sums of the real space correlations
$\langle \Delta^{\phantom{\dagger}}_\alpha(i+r)\Delta^\dagger_\alpha(i)\rangle$ with
standard conventions
$\Delta^\dagger_s(i) = c^\dagger_{i\uparrow} c^\dagger_{i\downarrow}$,
$\Delta^\dagger_d(i) = c^\dagger_{i\uparrow} 
\, \frac{1}{2} ( \, c^\dagger_{i+x\downarrow} 
- c^\dagger_{i+y\downarrow} 
+ c^\dagger_{i-x\downarrow} 
- c^\dagger_{i-y\downarrow} \,)$,
etc.~\cite{white1989attractive}.
These are shown in Fig.~\ref{fig:D_KE_Spairings}(d). 
Similar changes are observed at the crossover. A bifurcation in pairing with
 $p_x$ and $p_y$ symmetry, as well as the sharp change in $d$, and $s_{xx}$
 pair form factors, at $g_c \sim 0.9$ also signal the AF-BOW phase transition.
 An interesting, and intuitively reasonable, observation is that a BOW pattern 
formed along the $x$ or $y$-direction of the square lattice, increases pairing
 along plaquette diagonals ($p_{xy},p_{yx}$, and $s_{xx}$), but competes with
 pairing channels which are also aligned directly along the bonds ($d, p_{x}, p_{y}$).

{\it Conclusions:} In this work, we used DQMC simulations to map out the phase
 diagram of the single orbital square lattice optical SSHH model.  Our work
 fills in the full two dimensional phase diagram in the plane of positive $U$
 and $g$, hitherto only investigated along the $U=0$ and $g=0$ axes. The phase
 diagram is characterized by BOW and AF phases. At larger electron-phonon
 coupling strength, the $x$-$y$ symmetry is spontaneously broken and the system
 develops a BOW with a $\left( \pi, \pi \right)$ order. Given the different
 broken symmetries in the BOW and AF phases, and the sharp increase of BOW 
structure factor, the results indicate a first-order transition between 
these two phases. The most salient feature is that the ground state phase 
transition is much more sensitive to changes in electron-phonon coupling compared
 to variations in the Coulomb repulsion.
We interpret this as the result
of the lack of a direct competition between the two ordered phases. In the Hubbard-Holstein model, $U$ suppresses double occupancy while the
Holstein $g$ enhances it.  Thus the two interactions always conflict: they want the most fundamental structure, the site
occupations, to behave completely differently yielding $U_c \sim g^2$ (at $\omega_0=1$).  No such competition appears in the SSHH model.  Indeed, both interactions individually give rise to AF order leading to somewhat cooperative tendencies.
We thus argue that  this is why adding $U$ does not
significantly inhibit the formation of the BOW phase by the SSH phonons, leading to a near vertical phase boundary. 

In the AF region, for small electron-phonon coupling $g$, all the quantities
 that we analyzed, e.g.~the AF structure factor, kinetic energy, phonon displacement
 and double occupancy, remain approximately constant. For $g\gtrsim 0.2$, the double occupancy and the magnitude of the
 kinetic energy start increasing, while the AF structure factor decreases. This 
occurs even though the system still possesses true long range AF order as demonstrated
 by a finite size scaling analysis~\cite{supp}. This is due to the fact that the
 Hubbard AF and the SSH AF mechanisms are different~\cite{CaiYao2021, GoetzAssaad2021}.
 This new insight into the physics of the SSH-Hubbard Hamiltonian can be thought
 of as analogous to the well-established crossover from
Slater insulator to Mott-Hubbard insulator and from itinerant AF to Heisenberg
 AF with increasing $U$ in the Hubbard model ($g=0$)
 \cite{white89a,pruschke2003slater,raczkowski2021local}.
 We focused here on intermediate to strong coupling,, i.e. $U$ exceeding half the bandwidth $W=8t$ and $\omega_0=t$. Further investigation of the effect of $\omega_0$ on the cross-over is of interest.

Individually, the Hubbard and SSH Hamiltonians exhibit a rich panoply of
 phenomena when doped away from half-filling. The interaction $U$ leads to
 a complex mixture of pseudogap physics, strange metal behavior, stripe order
 and $d$-wave pairing when doped. The SSH model hosts polarons in the dilute
 limit which can bind to bipolarons and condense into superconducting phases.
 New phases of matter are thus likely to emerge from the study of regimes of the
 SSHH Hamiltonian away from half-filling. Work in this direction has already
 began as shown in~\cite{ZhangSvitsunov2021}. 

{\it Acknowledgments:} We acknowledge fruitful discussions with W.-T. Chiu and
 B. Cohen-Stead. CHF and RTS
  are supported by the grant DOE DE‐SC0014671 funded by the U.S.
 Department of Energy, Office of Science.
  D.P. acknowledges support from the Singapore Ministry of Education,
 Singapore Academic Research Fund
  Tier-II (Project No. MOE2018-T2-2-142). The computational work for this
 Letter was performed on
  resources of the National Supercomputing Centre, Singapore (NSCC) \cite{nscc}.\\ 

\bibliography{SSHH_Draft}

\renewcommand{\thefigure}{S\arabic{figure}}
\setcounter{figure}{0}
\renewcommand{\thesection}{S\arabic{figure}}
\setcounter{section}{0}
\renewcommand{\theequation}{S\arabic{equation}}
\setcounter{equation}{0}

\newpage
\centerline{\large \bf Supplemental Material}

\vskip0.20in
In these Supplemental Materials we provide details concerning
(a):
An approximate treatment of the SSHH model in terms of a pure Hubbard Hamiltonian
with a renormalized $U_{\rm eff}$;
(b):
The BOW Structure Factor with $U=4, \, g=1.5$ ;  $U=4,\ 15$ at $g=1.0$ for $\beta=16, \, L=12$ system;
(c):
Scaling of the antiferromagnetic structure factor
and, finally,
(d):
Structure factors at several different temperatures.



\vskip0.10in \noindent
\underline{\it (a) Effective Hubbard Hamiltonian:}
In the Holstein model, where the phonons couple to the local charge
density $H_{\rm el-ph} = \lambda \sum_i \hat X_i n_i$, one can integrate
out the phonon degrees of freedom in the anti-adiabatic limit
(i.e. ignoring the phonon kinetic energy $\hat P^2/2M$ term).  The result is an on-site attraction $U_{\rm eff} = -\lambda^2/\omega_0^2$. As
a consequence, the physics of the Hubbard-Holstein model at weak
$\lambda$ can be qualitatively interpreted in terms of a reduced on-site
repulsion $U-\lambda^2/\omega_0^2$.

\begin{figure}[hbpt]
   \centering
    \includegraphics[width=\columnwidth]{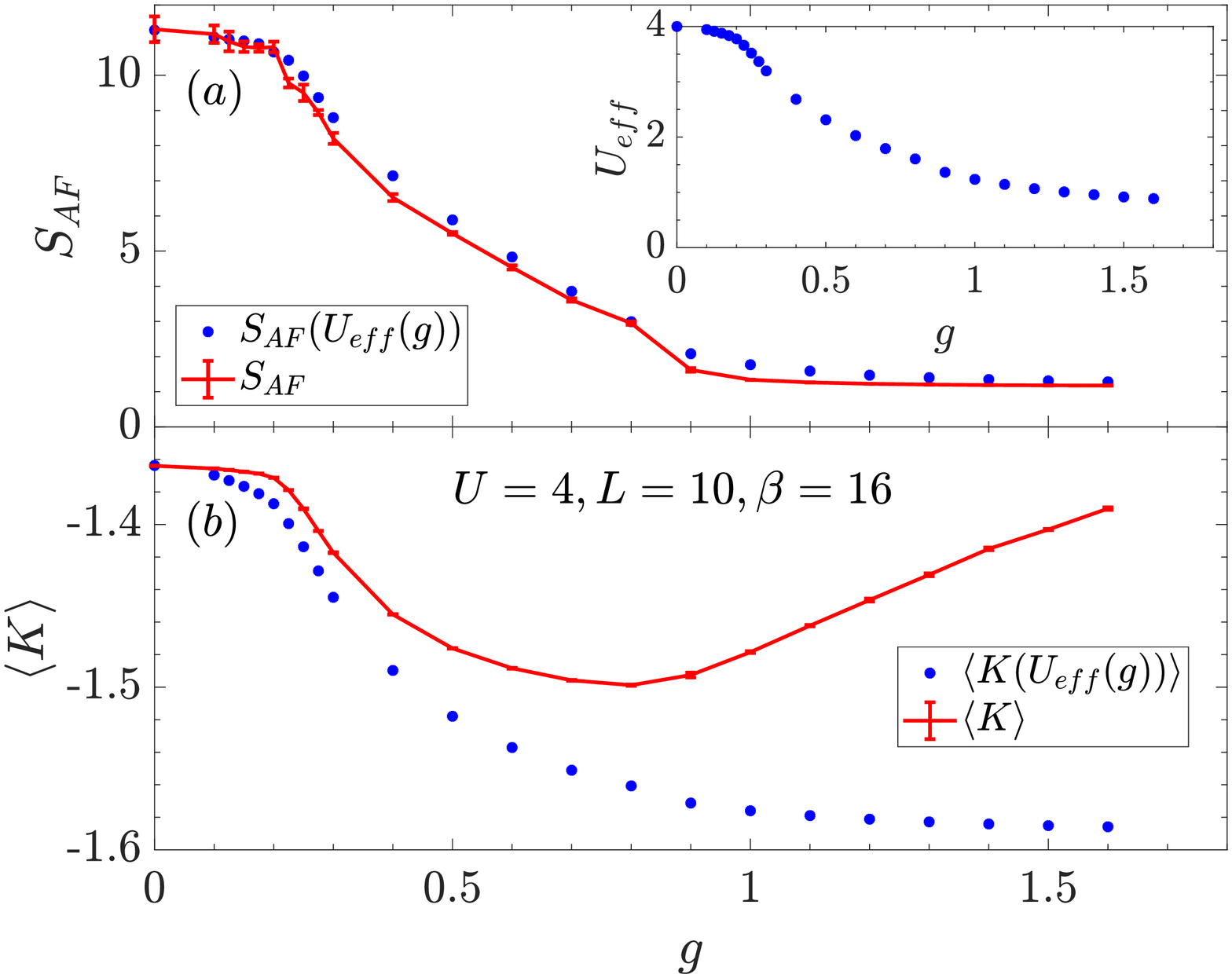}

\caption{Inset in panel (a): In SSHH model, when $U=4, \beta=16$ on a $L\times L =10 \times
10 $ square lattice, for each $g$, we define $U_{eff}$ as the onsite
electron repulsion in the pure Hubbard model which gives the same double
occupancy $D$.
(a) AF structure factor (b) Electron Kinetic energy given in SSHH model (red curve), for
$U=4$, varying $g$, $\beta=16$ on a $L\times L =10 \times 10$ square
lattice and in pure Hubbard model(blue dot) when $\beta=16$ on a
$L\times L =10 \times 10$ with $U_{\rm eff}$ defined in Fig.~\ref{fig: U_eff} (a) inset. Two curves are in reasonable agreement at relatively small $g$, while the discrepancy at large $g$ is because BOW phase can not be captured by the pure Hubbard model. }
\label{fig: U_eff}
\end{figure}

If one integrates out the SSH (bond) phonons on a two site dimer, the
resulting effective electron-only Hamiltonian has a renormalized $U$,
but also additional inter-site terms.  Nevertheless, one can still ask
the extent to which the SSHH model considered here can be
quantitatively modeled simply by a renormalized $U$.  We analyze this
issue as follows: We first take data for the double occupancy $D =
\langle n_{i\uparrow} n_{i\downarrow} \rangle$ both for the SSHH
model at fixed $U=4$ and varying $g$ and for the pure Hubbard model at
varying $U$.  For each $g$, We define $U_{\rm eff}(g)$ to be the value
of $U$ in the pure Hubbard case which gives the same $D$ for the
SSHH model.  The result is given in the inset of panel (a) in Fig.~\ref{fig: U_eff}.

Next we compare the values of {\it other} observables between
SSHH at $U=4$ and varying $g$ with pure Hubbard at $U_{\rm eff}$.
By construction, the values for $D$ match perfectly. 
Fig.~\ref{fig: U_eff}(a,b) gives the results for the 
antiferromagnetic structure factor $S_{\rm AF}$
and kinetic energy $\langle K \rangle$
respectively.  For $S_{\rm AF}$ the agreement between the actual SSHH
data and the effective model is remarkable- the results match to within a
few error bars across the complete range of $g$.  Perhaps not surprisingly, the SSHH model captures the more abrupt change in $S_{\rm AF}$ at the AF-BOW phase transition than the effective model, which has
no such transition.

The kinetic energy agreement is less good quantitatively, but still quite accurate qualitatively for small $g$. The effective model of course cannot capture the reduction in the magnitude of the kinetic energy at large $g$ which occurs upon entry into the BOW phase.  It is also observed in Fig.~\ref{fig: U_eff}(b) that the kinetic energy of the SSHH model remains more constant for small $g$ than does the kinetic energy of the effective model, an effect reminiscent of the appearance of weak $g$ AF regime discussed in the main text.
The effective model
gives a rough context in which to understand the suppression of magnetism by $g$.  
The resulting accuracy of panel (a) of Figure S1 is of additional interest:  it is not obvious that
adjusting $U$ to get a match for a local observable like double occupancy would also give a good match for intersite magnetic correlations.

 \begin{figure}[htp]
     \centering

     \includegraphics[width=1.0\columnwidth]{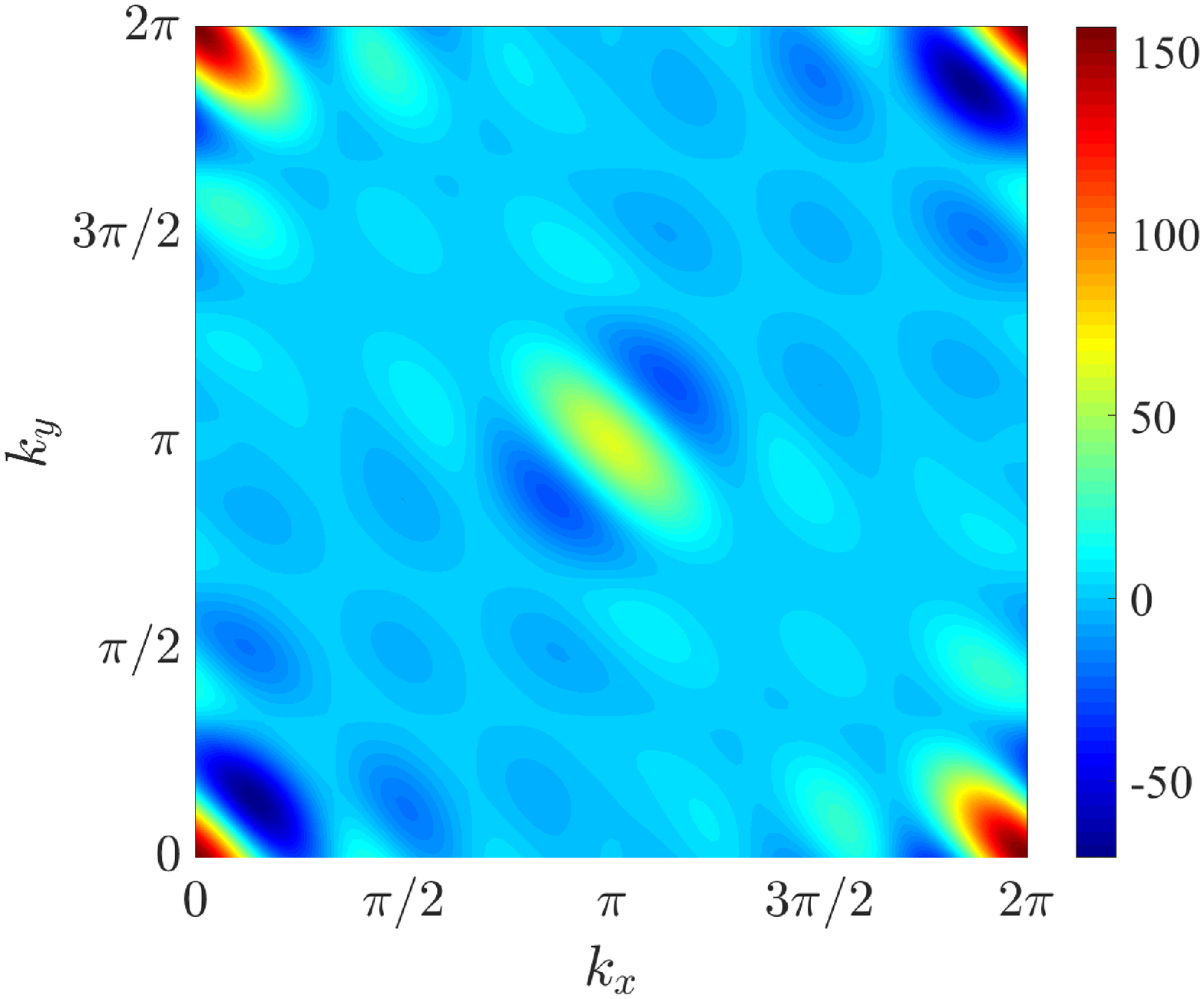}
     \caption{Bond ordered wave structure factor, $S_{K_{x}} \left( k_x, k_y \right)$ as a function of momentum $k_x$, $k_y$
 for $U=4, \,  g=1.5, \, \beta = 16, \, L = 12$.
     }
     \label{fig:SKx_kxky}
 \end{figure} 

\begin{figure}
    \centering
    \includegraphics[width=\columnwidth]{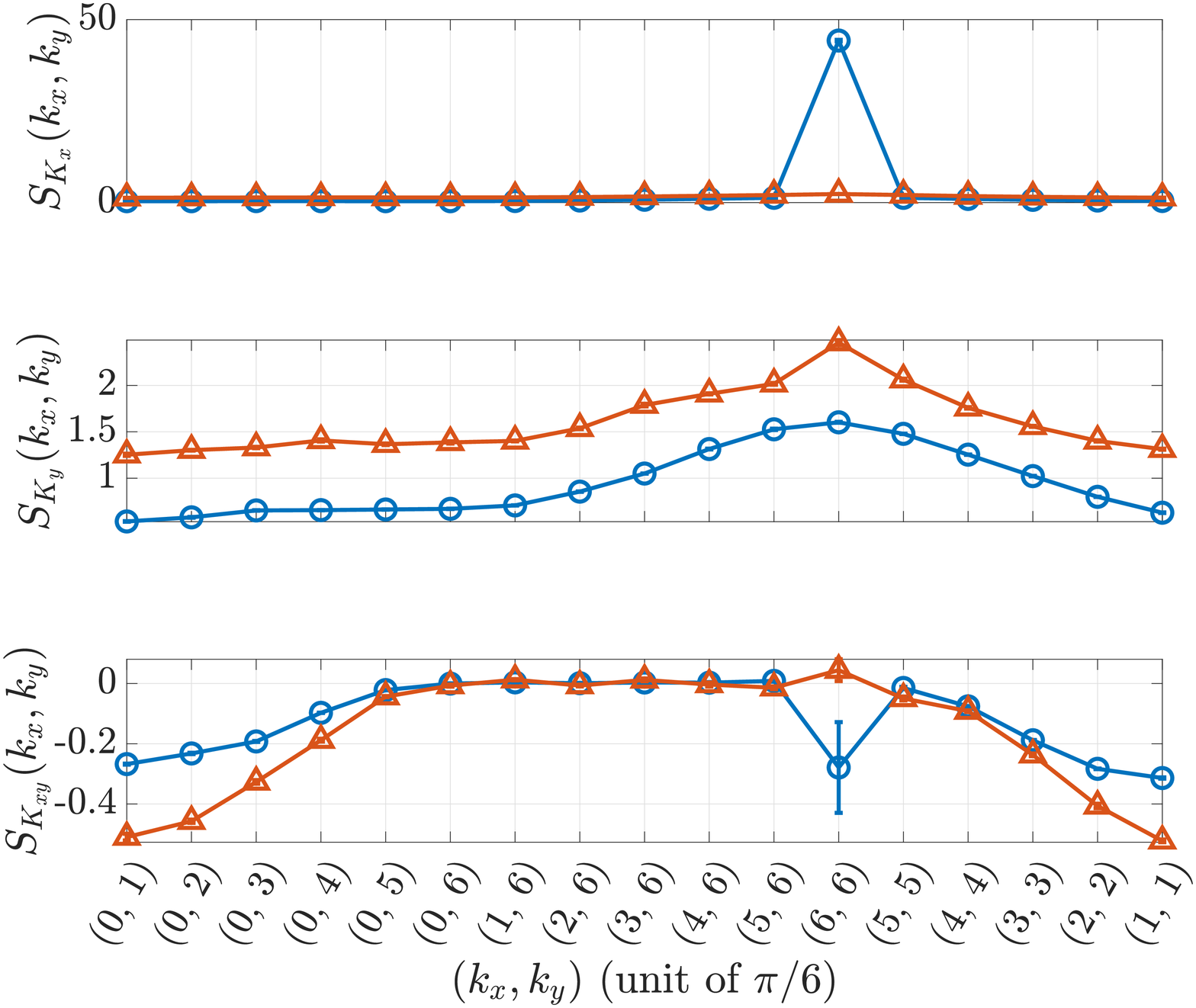}
    \caption{Bond ordered wave structure factor, $S_{K_x}$ (top), $S_{K_y}$ (center), $S_{K_{xy}}$ (bottom) for different $(k_x, k_y)$ momenta.
    The blue and red curves correspond to $U = 4$ and $U = 15$ respectively.
    The other system parameters  are $g = 1.0,\ \omega = 1$,\ $L = 12$, and $\beta=16$.}
    \label{fig:BOW_triangle_momenta}
\end{figure}

\vskip0.10in \noindent
\underline{\it (b) Bond ordered wave structure factor:}
For all simulations, we start with a phonon configuration that favors the development of XX bond ordered wave (BOW) pattern, i.e.~$\left( \pi, \pi \right)$ order of the $x$ bonds.
As noted in the main text, we have verified that for small lattices and high temperatures,
correlation functions are independent of initial configuration, and evolve to
a consistent long time state.  In addition, for all lattice sizes and
temperatures, BOW melts rapidly when it is not supported by the parameters of the simulation.  Thus our choice of starting configuration serves only to minimize long equilibration
times in the BOW phase, and does not affect our determination of the phase diagram.
Fig.~\ref{fig:SKx_kxky} shows the BOW structure factor $S_{K_x}(k_x,k_y)$, the Fourier transform of bond-bond kinetic energy correlations in real space, as a function of momenta $k_x$ and $k_y$. Besides a large value at $(k_x,k_y)=(0,0)$, which is actually the sum of all spatial correlations, a clear peak is observed at $(k_x,k_y)=(\pi,\pi)$ for $U=4, g=1.5, L=12, \beta=16$ system (in BOW phase).
In Fig.~\ref{fig:BOW_triangle_momenta}, we show the BOW structure factors for different directions and momenta when the system is in the BOW phase, i.e. $U = 4,\ g = 1.0$.
For all $S_{K_{x}}, S_{K_{y}}$ and $S_{K_{xy}}$, we find a peak only at $S_{K_{x}} (\pi, \pi)$.
It is worth pointing out that, despite the large error bar at $S_{K_{xy}} (\pi, \pi)$, the negligible magnitude of its structure factor suggests that there is no notable structure in the XY direction.
When the system is not in the BOW phase, i.e. $U = 15,\ g = 1.0$, the BOW structure factors for all three directions become negligible.


\begin{figure}[htp]
    \centering
    \includegraphics[width=1\columnwidth]{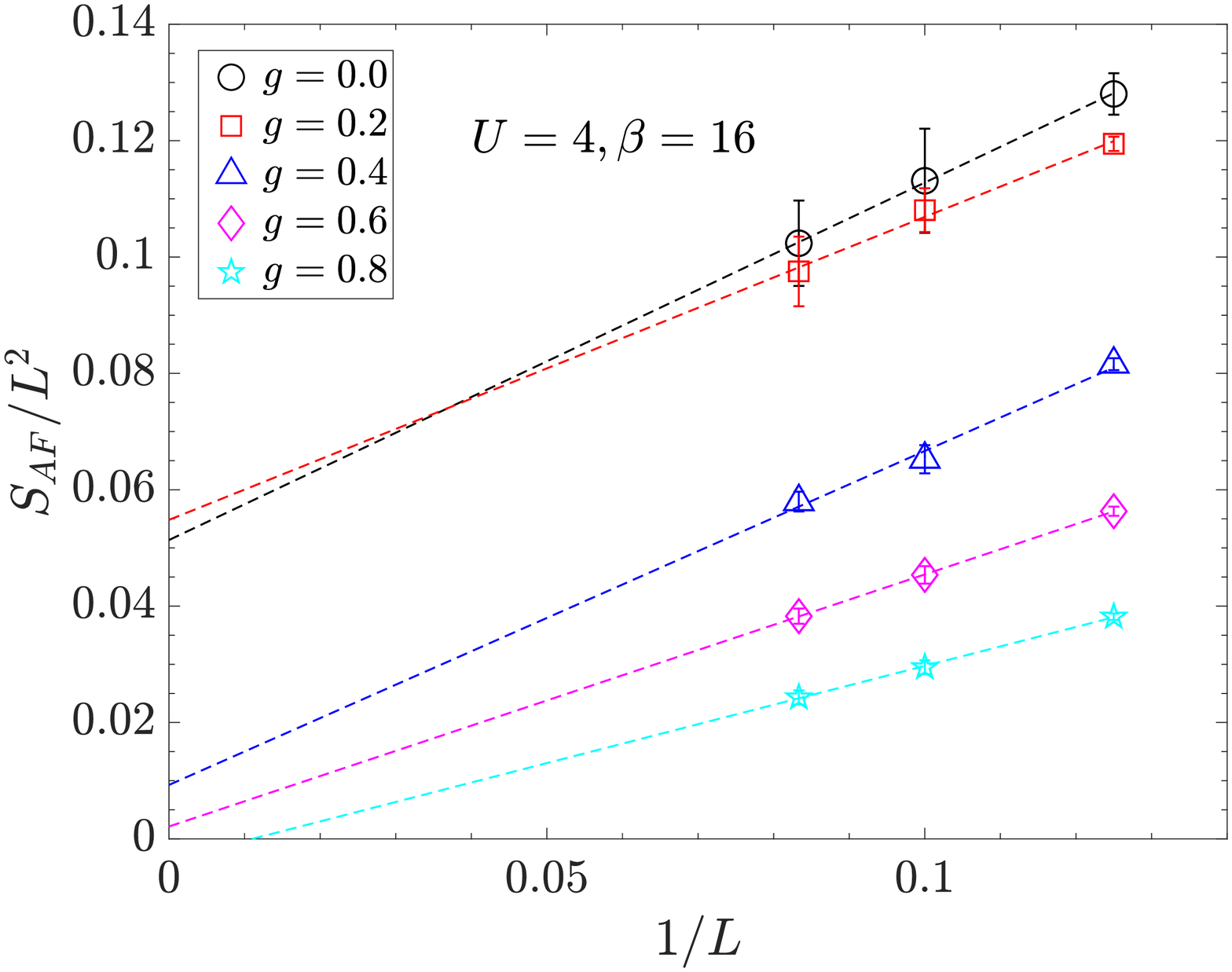}
    \caption{Finite size scaling of the antiferromagnetic structure factor, $S_{\mathrm{AF}}/L^{2}$, for $g = 0,\ 0.2,\ 0.4,\ 0.6,\ 0.8$.
    The dotted lines are extrapolated from the simulations data.
    For all results shown in this panel, $U = 4,\ \beta = 16$.
    }
    \label{fig:SafFSS}
\end{figure}

\begin{figure}[htp]
    \centering
    \includegraphics[width=1\columnwidth]{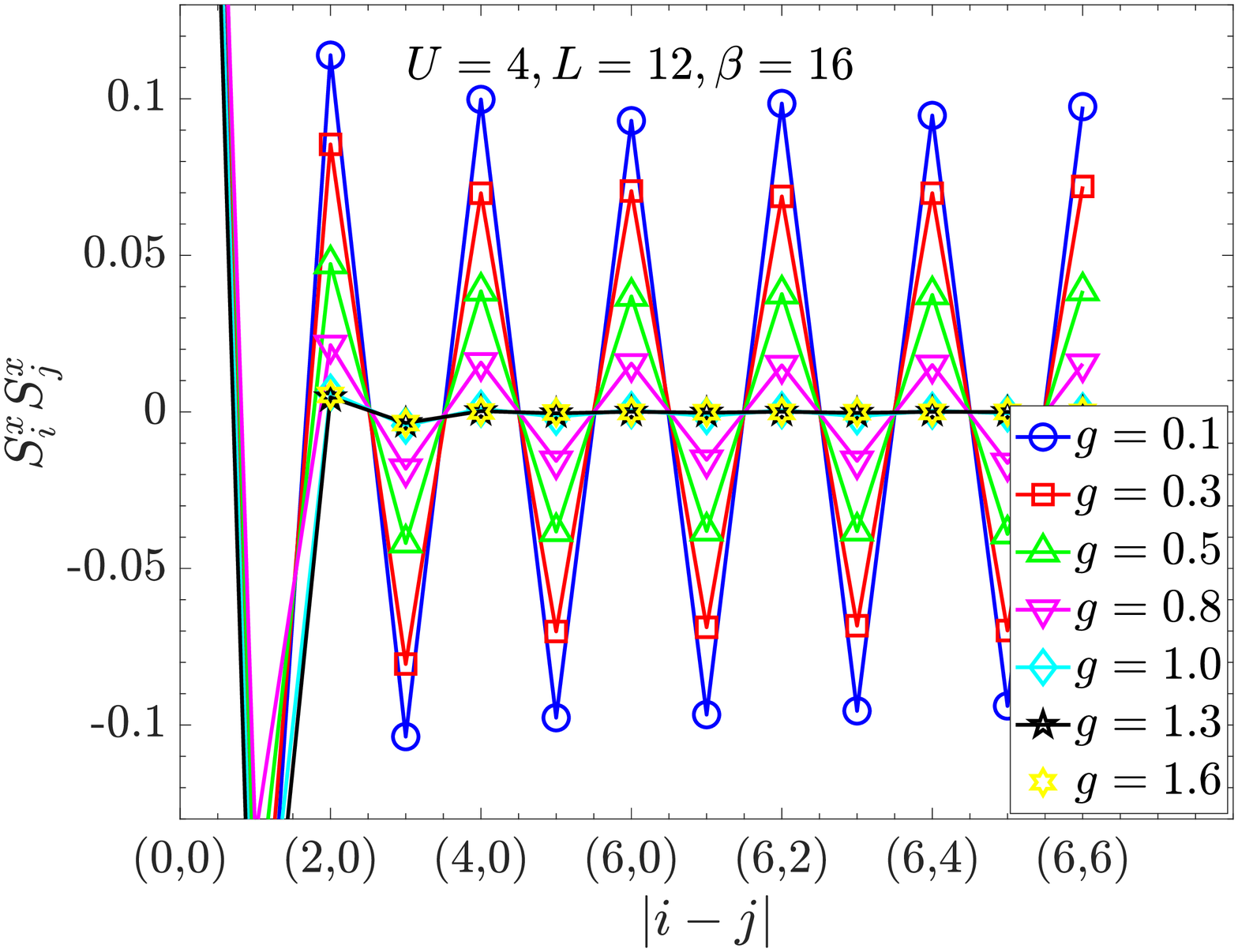}
    \caption{Spin-spin correlation, $\langle S_{i}^{x}S^{x}_{j}\rangle $ versus spatial distance $\left|i-j\right|$ for various $g$.
    }
    \label{fig:SafCF}
\end{figure}

\vskip0.10in \noindent
\underline{\it (c) Antiferromagnetic structure factor:}
In Fig~\ref{fig:SafFSS}, we perform finite size scaling (FSS) of the antiferromagnetic structure factor, $S_{\mathrm{AF}}/L^{2}$ and show that the antiferromagnetic (AF) structure is present in the thermodynamic limit for both AF regions.
At $U = 4,\ g = 0$, the system is in the AF phase.
As $g$ increases, $S_{\mathrm{AF}}/L^{2}$ decreases in the thermodynamic limit.
However, it is still finite and therefore indicative of the AF phase.
At $g = 0.8$, we notice that FSS of $S_{\mathrm{AF}}/L^{2}$ becomes less than zero.
At the same time, the BOW structure factor, $S_{K_{x}} \left( \pi, \pi \right)$, rises sharply to a finite value, indicating the direct transition from the AF phase to the BOW phase.

In Fig.~\ref{fig:SafCF}, the AF correlation function in real space gives similar conclusions.
The oscillation of the AF correlation function, which does not decay with the spatial distance, gives evidence to the presence of the AF phase.
As $g$ increases, the magnitude of this oscillation decreases.
The strong oscillation of the AF correlation function for $g < 0.8$, and the absence of BOW correlation, show that in this parameter region the system is AF.
\begin{figure}[htp]
    \centering
    \includegraphics[width=1\columnwidth]{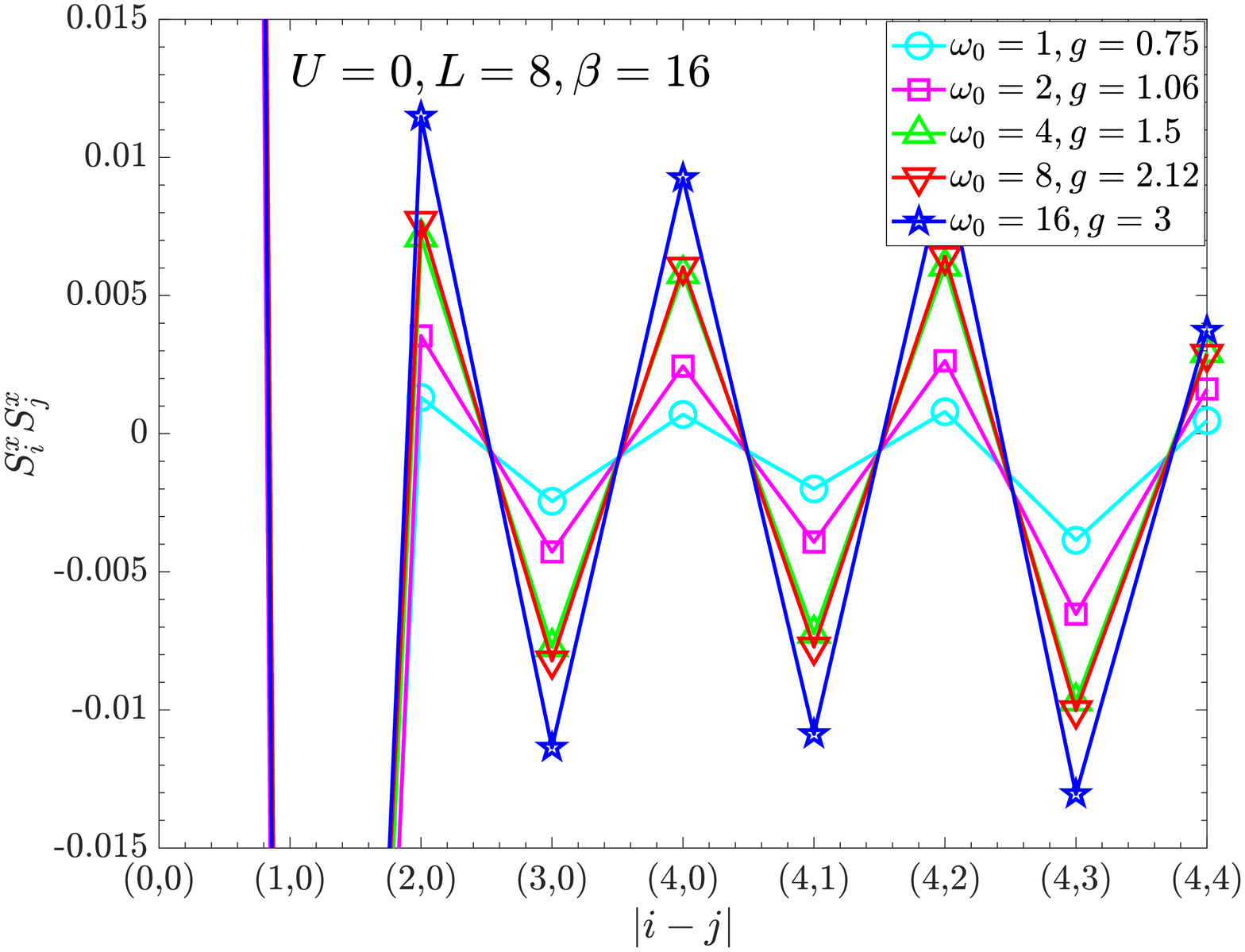}
    \caption{Spin-spin correlation, $\langle S_{i}^{x}S^{x}_{j}\rangle $ versus
    spatial distance $\left|i-j\right|$ for several phonon frequencies, $\omega_0$, for the pure SSH model on an $8\times8$ square lattice.  Increasing $\omega_0$ at fixed $g^2/\omega_0$ enhances AF order.
    }
    \label{fig:SxSx_U0L8beta16}
\end{figure}
\begin{figure}[htb]
    \centering
    \includegraphics[width=1\columnwidth]{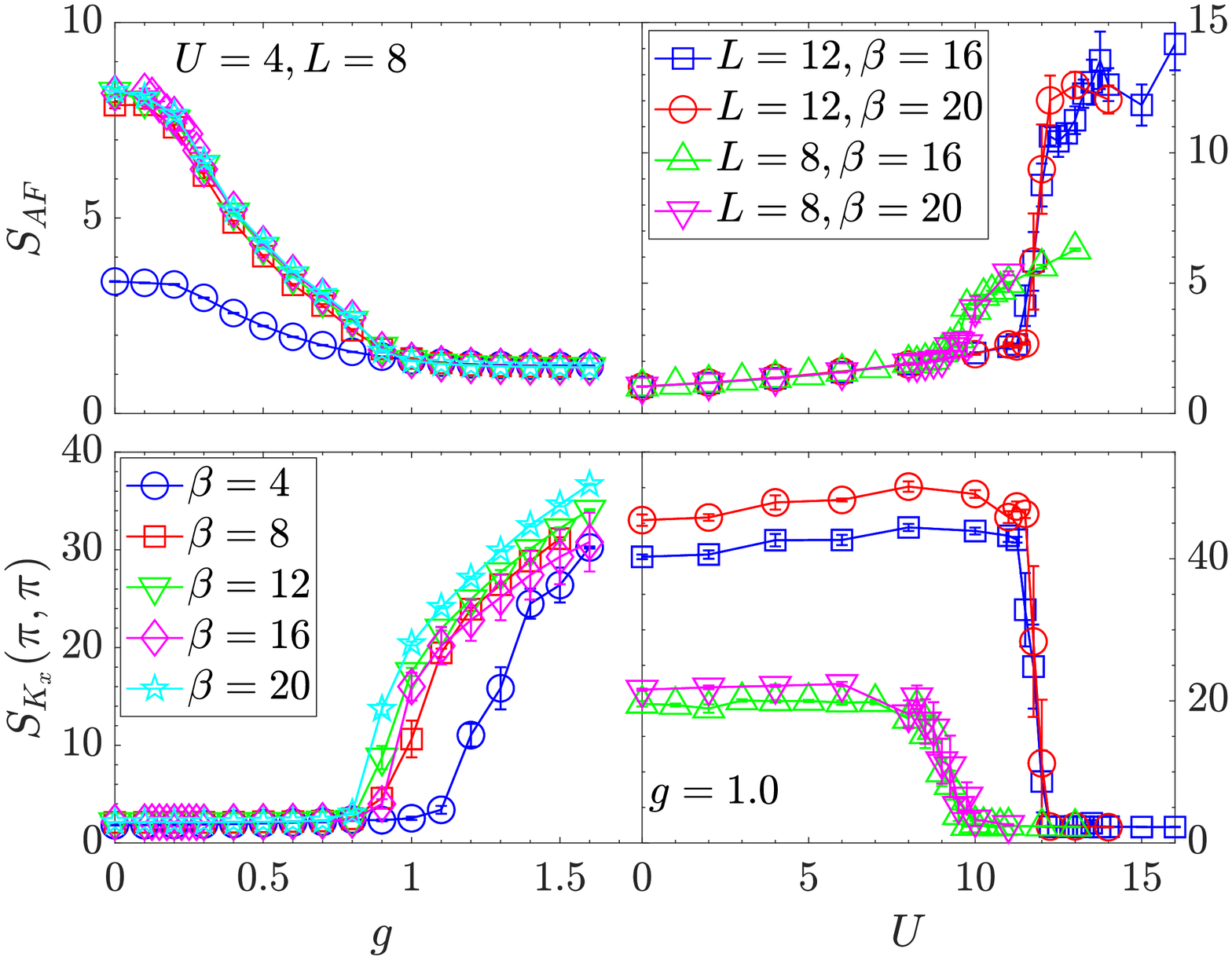}
    \caption{Left panels: AF and BOW structure factors vs. electron-phonon coupling $g$ for
 a fixed $U=4$, at several different temperatures $\beta=4,8,12,16,20$. Right panels:  AF and BOW structure factors
 vs.~onsite Coulomb interaction $U$ for a fixed $g=1.0$, $\beta=16,20$ on $8 \times 8$ and $ 12 \times 12$
 lattices. Both panels indicate $\beta=16$ is low enough to capture the ground state physics for the SSHH model.
    }
    \label{fig:SF_diff_low_beta}
\end{figure}

We also verify that for the pure SSH model($U=0$), when $g \lesssim g_c$, long range AF order exists at
low temperature ($\beta=16$).  The spin-spin correlation as a function of lattice site separation
 is shown in Fig.~\ref{fig:SxSx_U0L8beta16}. As $\omega_0$ increases with $g^2/\omega_0$ (which is proportional
 to the spin exchange strength in the anti-adiabatic limit) fixed, the AF order is strengthened, which
 is consistent with the conclusion in \cite{CaiYao2021,GoetzAssaad2021}


\vskip0.10in \noindent
\underline{\it (d) Structure factors at several different temperatures}

In the left panels of Fig.~\ref{fig:SF_diff_low_beta}, the AF and BOW structure factors are plotted as functions of
electron-phonon coupling strength $g$ for a fixed $U=4$ on a $8 \times 8$ square lattice. The curves almost
 overlap when $\beta \gtrsim 8$, indicating the temperature we use ($\beta=16$) in the main text is low enough
 to capture the ground state physics. Similarly, the right panels are structure factors $S_{AF}$ and
 $S_{K_x}(\pi, \pi)$ vs.~$U$. Although their magnitudes at $\beta=20$ in the
 ordered phase are slightly larger than those given by $\beta=16$, the transition point given by
 them coincide for both $L=8$ and $L=12$ lattices respectively.  There is
 some finite size effect going from $L=8$ to $L=12$ in both Fig.~\ref{fig:SF_diff_low_beta} and
 Fig.~\ref{fig:PD} but there is no such effect
 between $L=10$ and $L=12$.


\vskip0.20in


\end{document}